\documentclass[aps,prc,twocolumn,superscriptaddress]{revtex4-1}
\usepackage{CJK}
\usepackage{graphicx}
\usepackage{hyperref}
\usepackage{amssymb}
\usepackage{amsmath}
\usepackage[normalem]{ulem}
\usepackage{url}
\usepackage{color}

\begin{document}
\begin{CJK*} {UTF8}{} 

\title{Insights on pion production mechanism and symmetry energy at high density}

\author{Yangyang Liu}
\affiliation{China Institute of Atomic Energy, Beijing 102413, China}
\author{Yongjia Wang}
\affiliation{School of Science, Huzhou University, Huzhou 313000, China}
\author{Ying Cui}
\affiliation{China Institute of Atomic Energy, Beijing 102413, China}
\author{Cheng-Jun Xia}
\affiliation{School of Information Science and Engineering, Zhejiang University Ningbo Institute of Technology, Ningbo 315100, China}
\author{Zhuxia Li}
\affiliation{China Institute of Atomic Energy, Beijing 102413, China}
\author{Yongjing Chen}
\affiliation{China Institute of Atomic Energy, Beijing 102413, China}
\author{Qingfeng Li}
\email{liqf@zjhu.edu.cn}
\affiliation{School of Science, Huzhou University, Huzhou 313000, China}
\affiliation{Institute of Modern Physics, Chinese Academy of Sciences, Lanzhou 730000, China}
\author{Yingxun Zhang}
\email{zhyx@ciae.ac.cn}
\affiliation{China Institute of Atomic Energy, Beijing 102413, China}
\affiliation{Guangxi Key Laboratory of Nuclear Physics and Technology, Guangxi Normal University, Guilin, 541004, China}







\date{\today}

\begin{abstract}
The $N\Delta\to NN$ cross sections, which take into account the $\Delta$-mass dependence of M-matrix and momentum $p_{N\Delta}$, are applied on the calculation of pion production within the framework of the UrQMD model. Our study shows that UrQMD calculations with the $\Delta$-mass dependent $N\Delta\to NN$ cross sections enhance the pion multiplicities and decrease the $\pi^-/\pi^+$ ratios. By analyzing the time evolution of the pion production rate and the density in the overlapped region for Au+Au at the beam energy of 0.4A GeV, we find that the pion multiplicity probes the symmetry energy in the region of 1-2 times normal density. The process of pion production in the reaction is tracked including the loops of $NN\leftrightarrow N\Delta$ and $\Delta\leftrightarrow N\pi$, our calculations show that the sensitivity of $\pi^-/\pi^+$ to symmetry energy is weakened after 4-5 N-$\Delta$-$\pi$ loops in the pion production path, while the $\pi^{-}/\pi^{+}$ ratio in reactions at near threshold energies remains its sensitivity to the symmetry energy. By comparing the calculations to the FOPI data, we obtain a model dependent conclusion on the symmetry energy and the symmetry energy at two times normal density is $S(2\rho_0)$=38-73 MeV within $1\sigma$ uncertainties. Under the constraints of tidal deformability and maximum mass of neutron star, the symmetry energy at two times normal density is reduced to $48-58$ MeV and slope of symmetry energy $L=54-81$ MeV, and it is consistent with the constraints from ASY-EOS flow data.
\end{abstract}


\maketitle
\end{CJK*}

\section{Introduction}
The isospin asymmetric nuclear equation of state is very important for understanding the objective for both nuclear physics and astrophysics. Recently, the values of neutron star masses, radii, and tidal deformability which are obtained from the binary neutron star merging event GW170817 attracted lots of analysis on their favored nucleonic equation of state at suprasaturation density~\cite{Abbott17,Abbott18,Annala18,Fattoyev18,Abbott19,Malik19,NBZhang19,WJXie19,CYTsang19,MBTsang19}, and the inferred symmetry energy at two times normal density being $39-53$ MeV~\cite{NBZhang19,WJXie19,HTong19}. In laboratory, the intermediate energy heavy ion collisions (HICs) can also provide the constraints of symmetry energy around twice the saturation-density by using the pion production ratios, and it becomes an important goal of nuclear scientific research~\cite{MBTsang19,Carlson17,NuPECC}.

The pions are mainly produced through $\Delta$ resonance decay in intermediate energy HICs, thus, the ratio of pion's multiplicity, i.e., $M(\pi^-)/M(\pi^+)$ (simply named as $\pi^-/\pi^+$ ratio), was supposed as a probe to constrain the symmetry energy at suprasaturation density~\cite{BALi02PRL,BAli2002NPA}. In 2007, FOPI published the pion data, such as $M(\pi)$ and $\pi^-/\pi^+$~\cite{Reisd07}, at the beam energy ranging from 0.4A to 1.5A GeV. Many theoretical calculations have been performed to extract the symmetry energy by best fitting the FOPI data of $M(\pi)$ and $\pi^-/\pi^+$. Those calculations clearly show that the $\pi^-/\pi^+$ ratio near the threshold energy is sensitive to the density dependence of the symmetry energy, but different conclusions on the constraints of density dependence of the symmetry energy at suprasaturation density have been made~\cite{Xiao2009PRLvesoft,Xie2013PLBsoft,FengZQ2010PLBsstiff,SK2015PRC}. It stimulates the hard works on both theoretical and experimental studies for deeply understanding the pion production mechanism. In the experimental study, the remeasurement of subthreshold pion production at MSU and RIKEN~\cite{spirit} for Sn+Sn at the beam energy of 270A MeV has been performed, and the data will come soon. In the theoretical side, for deeply understanding the pion production mechanism, there are lots of effort to explore it from the threshold effects of $NN\leftrightarrow N\Delta$~\cite{SK2015PRC,Ferini05}, pion potential~\cite{JXu13,WMGuo15,Cozma2017PRC,JH2014PRC,liu2018PRC,ZZhang17}, $\Delta$ potential\cite{BALi15,Cozma2016PLB}, cluster formation~\cite{Ikeno16}, Pauli blocking~\cite{Ikeno19}, energy conservation issue~\cite{Cozma2016PLB,ZZhang2018PRC}, and so on. The calculations show that including the different physics in transport models could influence the prediction of $\pi^-/\pi^+$~\cite{ZZhang17}. 

On the other hand, the model dependence should be well understood. The code comparison projects have been inspired in the last 20 years~\cite{Kolomeitsev05,Xujun16,YXZhang18,Akira19} for improving the reliability of transport models. Recently, the results of code comparison by Akira \textit{et al}.~\cite{Akira19} found that the better method on treating the baryon production and decay in the collision part of transport models is to adopt the time-step-free method, which automatically determines the order of two-body collision or resonance decay according to their collision time and decay time. This method has been adopted in many codes, such as UrQMD~\cite{Bass1998,QFL2005PRC,QFL2011}, JAM~\cite{NaraJAM1999,Ikeno16}, and SMASH~\cite{SMASH2016Weil}. Since the UrQMD model has been well designed for solving the particle's production and decay in the collision part of the transport equation, we decide to adopt it to investigate the pion production mechanism near the threshold energy and make a discussion on the symmetry energy by comparing the calculations with the FOPI data.

This paper is organized as follows: we first briefly introduce the UrQMD model, and then describe the interaction parameters and the cross sections we used. Considering there is still lack of the studies of the influence of $N\Delta\to NN$ cross sections on the yield of pions and the $\pi^{-}/\pi^{+}$ ratio in heavy ion collisions at intermediate energies, we consider a $\Delta$-mass dependent cross sections of the channel $N\Delta \rightarrow NN$ which is obtained based on the one boson exchange model, and investigate its effects on the $M(\pi)$ and $\pi^-/\pi^+$ near the threshold energy. The results on pion production mechanism, pion multiplicities and pion ratios obtained with UrQMD model are presented and discussed. Finally, we discussed the symmetry energy constraints from $\pi^-/\pi^+$ ratios, the tidal deformability, and maximum mass of neutron star.

\section{UrQMD model and $N\Delta\to NN$ cross sections}
The UrQMD model is a microscopic many-body approach to simulate the reaction of p-p, p-A, and A-A systems in the large energy range from SIS to LHC. It mainly consist of the initialization of projectile and target nuclei, the mean field and the collision term~\cite{Bass1998}.

In the UrQMD model, hadrons are represented by Gaussian wave packets with the width parameter $\sigma_r$. After the initialization of projectile and target nuclei, the time evolution of the coordinate and momentum of hadron $i$ is propagated according to the Hamilton equations of motion:

\begin{equation}
\label{eom}
\dot{\vec{r}}_{i}=\frac{\partial H}{\partial \vec{p}_{i}}, \, \dot{\vec{p}}_{i}=-\frac{\partial H}{\partial \vec{r}_{i}}.
\end{equation}
The Hamiltonian $H$ contains the kinetic energy and the effective interaction potential energy $U$~\cite{Zhang20FOP}.

The form of the isocalar part of potential energy density used in this work is,
\begin{eqnarray}
u &=& \frac{\alpha}{2}\frac{\rho^2}{\rho_0}+\frac{\beta}{\gamma+1}\frac{\rho^{\gamma+1}}{\rho_0^\gamma}\\\nonumber
&+&\frac{g_{sur,iso}}{2\rho_0}(\nabla \rho)^2+\frac{g_{sur,iso}}{2}(\nabla (\rho_n-\rho_p))^2\\\nonumber
&+& u_{md}.
\end{eqnarray}
The energy density related to the momentum dependent interaction is obtained based on the isospin independent momentum dependent interaction as in Ref.~\cite{Aichelin87}, i.e. $t_4\ln^2(1+t_5(\mathbf{p}_1-\mathbf{p}_2)^2)\delta(\mathbf{r}_1-\mathbf{r}_2)$, and it yields the effective mass $m^*/m=0.77$ ($m^*/m=(1+\frac{m}{p}\frac{dU}{dp})^{-1}$) at Fermi momentum. The parameter set we has used in the calculations is an updated SM EOS with $K_0=231$ MeV as in Table~\ref{tab:table1}. The calculations with SM EOS with $K_0=200$ MeV as in Ref.~\cite{YXYe18PRC} are also discussed in the following. 
\begin{table}[htbp]
\caption{\label{tab:table1}%
Parameters in UrQMD, $\alpha$, $\beta$ are in MeV, $g_{sur}$ and $g_{sur,iso}$ are in MeVfm$^2$. $t_4$ and $t_5$ are the coefficients in momentum dependent interaction, in MeV and MeV$^{-2}$, and $\rho_0=0.16\mathrm{fm}^{-3}$. Last two columns are $K_0$ in MeV and $m^*/m$.}
\begin{tabular}{lcccccccc}
\hline
\hline
$\alpha$ & $\beta$ & $\gamma$ & $g_{sur}$ & $g_{sur,iso}$ & $t_4$  & $t_5$ & $K_0$ & $m^*/m$\\
\hline
-221 & 153 & 1.31 & 19.5 & -11.3 & 1.57 & 5$\times$10$^{-4}$ & 231 & 0.77 \\
\hline
\hline
\end{tabular}
\end{table}

For the isovector part of potential energy, two forms of density dependence of symmetry potential energy density functional are adopted. One is the Skyrme-type polynomial form (form (a) in Eq.~(\ref{srho-lyy})) and another is the density power law form (form (b) in Eq.~(\ref{srho-lyy})). It reads,
\begin{eqnarray}
\label{srho-lyy}
 u_{sym}&=&S^{pot}_{sym}(\rho)\rho\delta^2\\\nonumber
 &=&\left\{
 \begin{array}{ll}
    ( A(\frac{\rho}{\rho_0})+B(\frac{\rho}{\rho_0})^{\gamma_s}+C(\frac{\rho}{\rho_0})^{5/3} )\rho\delta^{2}, & \mathbf{(a)}\\
    \frac{C_{s}}{2}(\frac{\rho}{\rho_{0}})^{\gamma_i}\rho\delta^2. & \mathbf{(b)}
  \end{array}
\right.
\end{eqnarray}
The parameters of Eq.~(\ref{srho-lyy}) used in this work are listed in Table.~\ref{tab:table2}, which correspond to five different density dependence of symmetry energy. The last four columns in Table.~\ref{tab:table2} are the corresponding values of symmetry energy coefficient $S_0=S(\rho_{0})$ and the slope $L_0=3\rho_{0}\left(\frac{\partial{S(\rho)}}{\partial\rho}\right)|_{\rho=\rho_{0}}$, with $S(\rho)=\frac{\hbar^2}{6m}(\frac{3\pi^2\rho}{2})^{2/3}+S^{pot}_{sym}(\rho)$, effective mass $m^*/m$ and symmetry energy at 2$\rho_0$, i.e. $S(2\rho_0)$.
\begin{table}[htbp]
\caption{\label{tab:table2}%
Parameters of symmetry potential in UrQMD. $A$, $B$, $C$, $C_s$, $S_0$ and $L$ are in MeV, $\gamma_s$ and $\gamma_i$ are dimensionless.}
\begin{tabular}{lcccccccc}
\hline
\hline
$\mathbf{S(\rho)_{a}}$ & $A$ & $B$ & $C$ & $\gamma_s$ & $S_{0}$ &  $L$ & $S(2\rho_0)$\\
\hline
$S_{1}$ & 62.84 & -38.30 & -6.39 & 1.1667 & $30.$ &  $46$  & 38.0 \\
$S_{2}$ & 20.37 & 10.75 & -9.28 & 1.3 & $34.$ &  $81$  & 57.3 \\
$S_{3}$ & 22.16 & -14.3 & 13.8 & 1.25 & $33.$ & $104$  & 73.5 \\
\hline
$\mathbf{S(\rho)_{b}}$ & $\frac{C_{s}}{2}$ & $\gamma_i$ & &  & $S_{0}$  & $L$ & $S(2\rho_0)$  \\
\hline
$G_{05}$ & $20$ & $0.5$ & - & - & $32.5$ &  $54$ &  47.7 \\
$G_{20}$ & $20$ & $2.0$ & - & - & $32.5$ &  $144$ &  99.5\\
\hline
\hline
\end{tabular}
\end{table}

The symmetry potential of ${\Delta}$ resonance is calculated from the symmetry potential of nucleon according to,
\begin{eqnarray}
\label{eq:VDelta}
V_{sym}^{\Delta^{++} } &=& V_{sym}^{p },\\\nonumber
V_{sym}^{\Delta^+ } &=& \frac{1}{3}V_{sym}^{n }+\frac{2}{3}V_{sym}^{p },\\\nonumber
V_{sym}^{\Delta^0 } &=& \frac{2}{3}V_{sym}^{n }+\frac{1}{3}V_{sym}^{p },\\\nonumber
V_{sym}^{\Delta^- } &=& V_{sym}^{n },
\end{eqnarray}
which is the same as those used in Refs.~\cite{BAli2002NPA,FengZQ2010PLBsstiff,Ferini05,SK2015PRC,Cozma2016PLB,QFL2005PRC}. The threshold effect and the pion optical potential have been studied with various models, but still with some puzzling inconsistency~\cite{JXu13,WMGuo15,Cozma2017PRC,ZZhang17}. Further studies are certainly required~\cite{liu2018PRC}. For simplicity, we will not consider the threshold effect in the present work, since introducing this effect needs to largely improve the collision treatments in UrQMD. We plan to fix it in the near future work. 

In the collision term, the medium modified nucleon-nucleon elastic cross sections are used as same as that in our previous works~\cite{YJWang18}. For the $NN\to N\Delta$ cross sections and decay width of $\Delta$, we use the standard values of UrQMD where the cross section of $NN\rightarrow N\Delta$ is obtained by fitting the CERN experiment data~\citep*{Bass1998,CERN1984} and decay widths depend on the mass of excited resonance~\cite{Bass1998}.

Concerning the $N\Delta\to NN$ cross sections, the popular way to obtain the $N\Delta\to NN$ cross sections is from the measured cross section of $NN\to N\Delta$ by using the detailed balance, where the treatment of the $\Delta$ mass dependence is partly considered by using the proposed method in Ref.~\cite{Daniel1991}. Here, we apply the $\Delta$ mass dependent $N\Delta\to NN$ cross sections which were recently calculated based on the one-boson-exchange model (OBEM)~\cite{YCui2019}, i.e., $\sigma_{N\Delta\to NN}^{OBEM}(\sqrt s, m_\Delta)$. At the given value of $m_\Delta$, it is calculated as,
\begin{eqnarray}
\label{eq:xsndnnM}
&&\sigma^{OBEM}_{N\Delta\to NN}(\sqrt s, m_\Delta)\\\nonumber
&=&\frac{1}{1+\delta_{N_{1}N_{2}}}\frac{1}{64\pi^2}\int \frac{|\textbf{p}'_{\text{12}}|}{\sqrt{s_{\text{34}}}\sqrt{s_{\text{12}}}|\textbf{p}'_{\text{34}}(m_\Delta)|}\\\nonumber
&&\times \overline{|\mathcal{M}_{N\Delta(m_\Delta) \to NN}|^2}  d\Omega,
\end{eqnarray}
$\textbf{p}'_{\text{34}}$ is the momentum of incoming $N$ or $\Delta$, and $\textbf{p}'_{\text{12}}$ is the momentum of outgoing $N$ in center of mass frame. For M-matrix, there is
\begin{eqnarray}
&&\overline{|\mathcal{M}_{N\Delta(m_\Delta) \to NN}|^2}\\\nonumber
&&=\frac{(2s_{1}+1)(2s_2+1)}{(2s_{3}+1)(2s_4+1)}\overline{|\mathcal{M}(m_\Delta)|^2},
\end{eqnarray}
at the same $\Delta$ mass for both process. $s_i$ is the spin of particle $i$, and $\overline{|\mathcal{M}(m_\Delta)|^2}$ is the M-matrix for $NN\to N\Delta$. The form of M-matrix in the calculation of $\sigma_{N\Delta\to NN}^{OBEM}(\sqrt s, m_\Delta)$ is as same as in Ref.~\cite{YCui2019}, and the parameters in M-matrix are determined by fitting the experimental data of $pp\to n\Delta^{++}$~\citep*{Bass1998,CERN1984}. Thus, a $\Delta$ mass dependencies of $\mathbf{p}'_{34}$ and M-matrix in the calculation of cross section of $N\Delta\to NN$ is taken into account. The different channels are determined based on the relationship $\sigma_{n\Delta^{++}\to pp}$ : $\sigma_{p\Delta^{-}\to nn}$ : $\sigma_{n\Delta^+\to np}$ : $\sigma_{p\Delta^0\to np}$: $\sigma_{n\Delta^0\to nn}$ : $\sigma_{p\Delta^+\to pp}$ = 3: 3: 2: 2: 1: 1. More details can be found in~\cite{YCui2019}. We incorporate the cross sections of $\sigma^{OBEM}_{N\Delta\to NN}(\sqrt s,m_{\Delta})$ directly into the UrQMD model.

\begin{figure}[htbp]
\centering
\includegraphics[angle=0,scale=0.30]{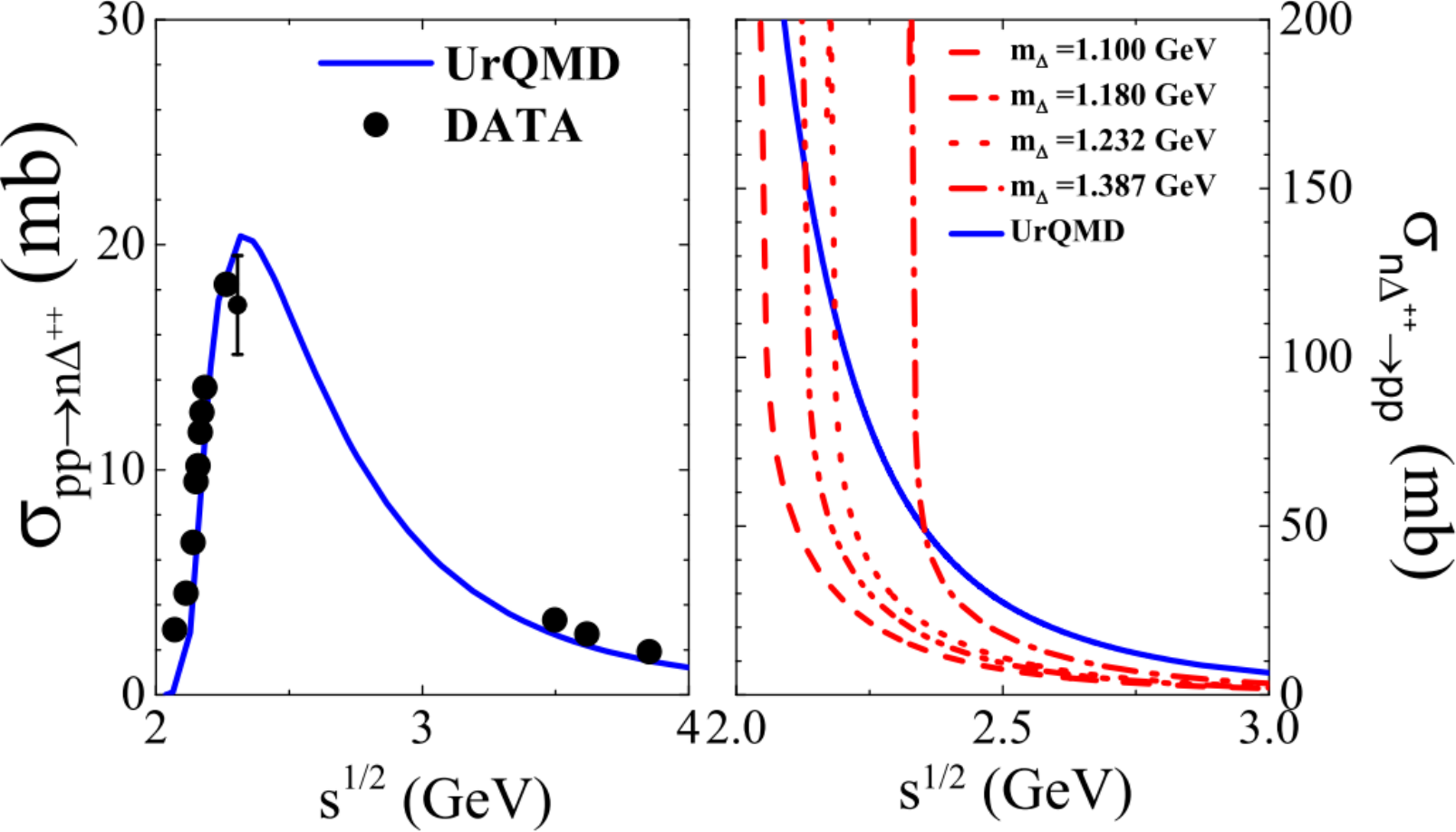}
\setlength{\abovecaptionskip}{0pt}
\vspace{2em}
\caption{(Color online) Left panel: the cross section of $pp\to n\Delta^{++}$ used in UrQMD, data is taken from~\cite{Bass1998,CERN1984}. Right panel: the cross sections of $n\Delta^{++} \to pp $ used in UrQMD and obtained from OBEM model.}
\setlength{\belowcaptionskip}{0pt}
\label{fig:sigma}
\end{figure}

In the left panel of Figure~\ref{fig:sigma}, we present the $\sigma_{pp\to n\Delta^{++}}$ used in the UrQMD~\cite{Bass1998,CERN1984}.
In the right panel of Figure~\ref{fig:sigma}, we present $\sigma_{n\Delta^{++}\to pp}^{UrQMD}(\sqrt s)$ (blue line) and $\sigma_{n\Delta^{++}\to pp}^{OBEM}(\sqrt s, m_\Delta)$ (red lines) at four values, i.e., $m_\Delta$=1.10, 1.18, 1.232, 1.387 GeV. Here, $\sigma_{n\Delta^{++}\to pp}^{UrQMD}(\sqrt s)$ is the cross sections used in previous UrQMD model calculations~\cite{QFL2006JPG32}, and the cross sections for other channels are determined based on the Clebsch-Gordan relationship. The calculations show that $\sigma_{n\Delta^{++}\to pp}^{OBEM}(\sqrt s, m_\Delta)$ is lower than $\sigma_{n\Delta^{++}\to pp}^{UrQMD}(\sqrt s)$ in low mass $\Delta$ cases, especially around the threshold energy. It means that there is too strong hard $\Delta$ absorption when one uses $\sigma_{n\Delta^{++}\to pp}^{UrQMD}(\sqrt s)$, and it results in the underestimation of the pion multiplicity which has been observed in our previous studies~\cite{QFL2006JPG32} (also can be noticed in Fig.6 of this paper). We note here that, as overall, the probability for a nucleon to undergo inelastic scattering and to become a $\Delta$ is less than 10\% in HICs around 1A GeV, e.g., see Fig.~\ref{fig:Ncoll-ratios} in the present work and Fig.2 in Ref.~\cite{bass1995}, thus the influence of $N\Delta\to NN$ on nucleonic observables are weak.


\section{Results and discussions}

\subsection{Pion production mechanism}
Before drawing the conclusion on the symmetry energy at high density, it is important to investigate the mechanism of pion production in the UrQMD model by analyzing the time evolution of the density in the compressed region, $\Delta$ production, pion production, and the collision/decay number of $NN\leftrightarrow N\Delta$, $N\pi\leftrightarrow \Delta$ for Au+Au at the beam energy from 0.4A GeV to 1.0A GeV.

\begin{figure}[htbp]
\centering
\includegraphics[angle=0,scale=0.30]{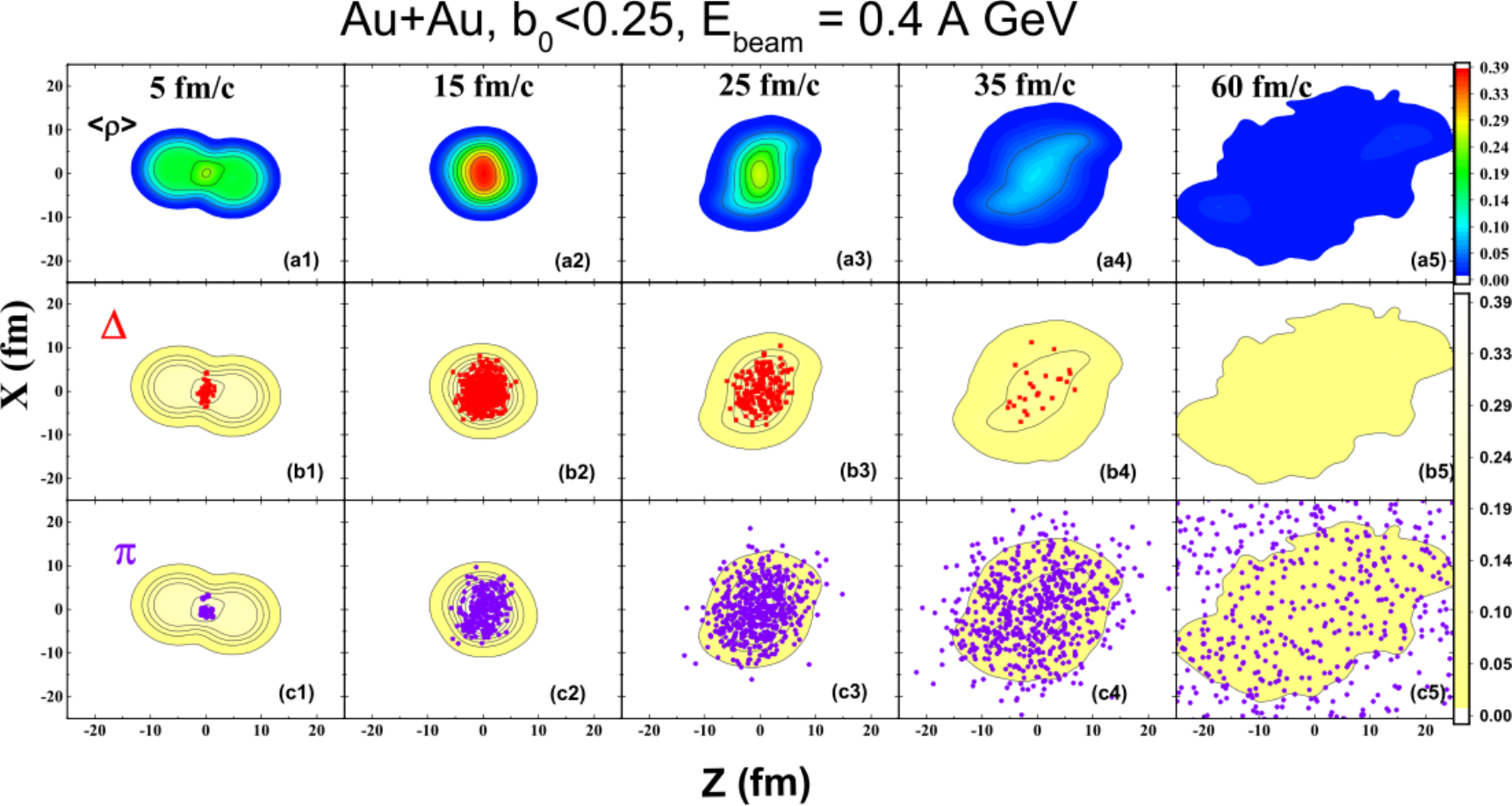}
\setlength{\abovecaptionskip}{0pt}
\vspace{2em}
\caption{(Color online) Panels (a1)-(a5): snap shots of density contour plots for central collisions ($b_0<0.25$), Au+Au reaction at a beam energy of 0.4A GeV. Panels (b1)-(b5) are the positions of $\Delta$ resonance (red symbols), and panels (c1)-(c5) are the positions of $\pi$ mesons (violet symbols). These figures are obtained from 100 events. }
\setlength{\belowcaptionskip}{-20pt}
\label{fig:contour}
\end{figure}

To obtain an intuitive view on the reaction process, the time evolution of the average density contour plots (upper panels), of the position of $\Delta$ (middle panels), and of the position of $\pi$ (bottom panels), for the central collision of $^{197}$Au+$^{197}$Au at 0.4A GeV, are presented in Figure~\ref{fig:contour}. The plots are obtained with 100 events. As shown in the upper panels of Figure~\ref{fig:contour}, the projectile and target start to hit at a large velocity around 5 fm/$c$ and few $\Delta$ resonances appear in the central region. It can be observed in the panel (b1) where the red points represent the position of $\Delta$. Around 15 fm/$c$, the density of the compressed system reaches the maximum, and lots of $\pi$ (violet points) are produced following the $\Delta$ production in the compressed region (see panels (a2), (b2), (c2)). As the time evolves to 25 fm/$c$, the number of $\Delta$ starts to decrease because the $\Delta$s decay to nucleon and pion. One can find that the number of $\pi$ become larger and larger with time in the bottom panels of Figure~\ref{fig:contour}. After 25 fm/$c$, the system expands to lower density, and $\Delta$s are mainly consumed by $\Delta\to N+\pi$ process. The produced $\pi$s propagate from high density of the compression phase to low density of the expansion phase, and during the time, $\pi$s may experience several $N-\Delta-\pi$ loops before freeze out. 

\begin{figure}[htbp]
\centering
\includegraphics[angle=0,scale=0.34]{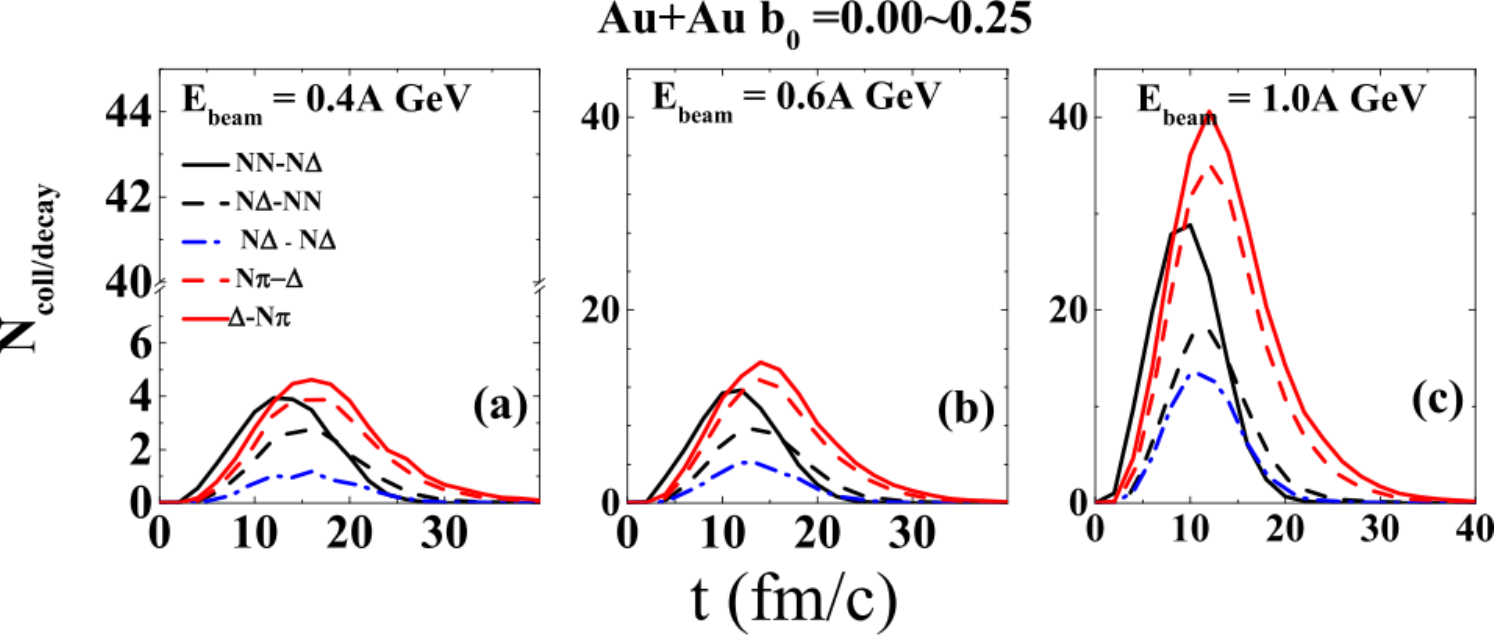}
\setlength{\abovecaptionskip}{0pt}
\vspace{2em}
\caption{(Color online) The number of the $\Delta$-related collision/decay as a function of time at beam energy $E_{beam}$=0.4A GeV (a), 0.6A GeV (b), and 1.0A GeV (c).}
\setlength{\belowcaptionskip}{0pt}
\label{fig:Ncoll-t}
\end{figure}

In detail, we present the collision/decay number of different channels, such as $N_{coll}(NN\to N\Delta)$, $N_{coll}(N\Delta \to NN)$, $N_{coll}(N\pi\to \Delta)$, and $N_{decay}(\Delta\to N\pi)$ as a function of time, in Figure~\ref{fig:Ncoll-t}. The $N_{coll}(N\Delta\to NN)$ is smaller and reaches the maximum later in time than $N_{coll}(NN\to N\Delta)$, because of the higher threshold energy for $N\Delta\to NN$ than $NN \to N\Delta$. For example, Au+Au at 0.4A GeV, the peak of the $N_{coll}(N\Delta\to NN)$ appears around 12.5 fm/$c$, while the $N_{coll}(N\Delta \to NN$), $N_{decay}(\Delta\to N\pi)$, $N_{coll}(N\pi\to \Delta)$, peak around 15 fm/$c$. One important feature is that the $N_{decay}(\Delta\to N\pi)$ and $N_{coll}(N\pi\to \Delta)$ are higher than $N_{coll}(NN\to N\Delta)$ and $N_{coll}(N\Delta\to NN)$ after 15 fm/$c$. It means the loop of $N\pi\leftrightarrow \Delta$ has a larger possibility than the loop of $NN\leftrightarrow N\Delta$ in the late stage of heavy ion collisions. Also, one can find that the loop of $N\pi\leftrightarrow \Delta$ lasts a long time, i.e., to 35 fm/$c$, for the beam energies we studied. Similar behaviors can be found for other beam energies. Our calculations imply that the $M(\pi)$ and $\pi^-/\pi^-$ contain the information of symmetry energy at a large region of density variation during the system evolution.

\begin{figure}[htbp]
\centering
\includegraphics[angle=0,scale=0.35]{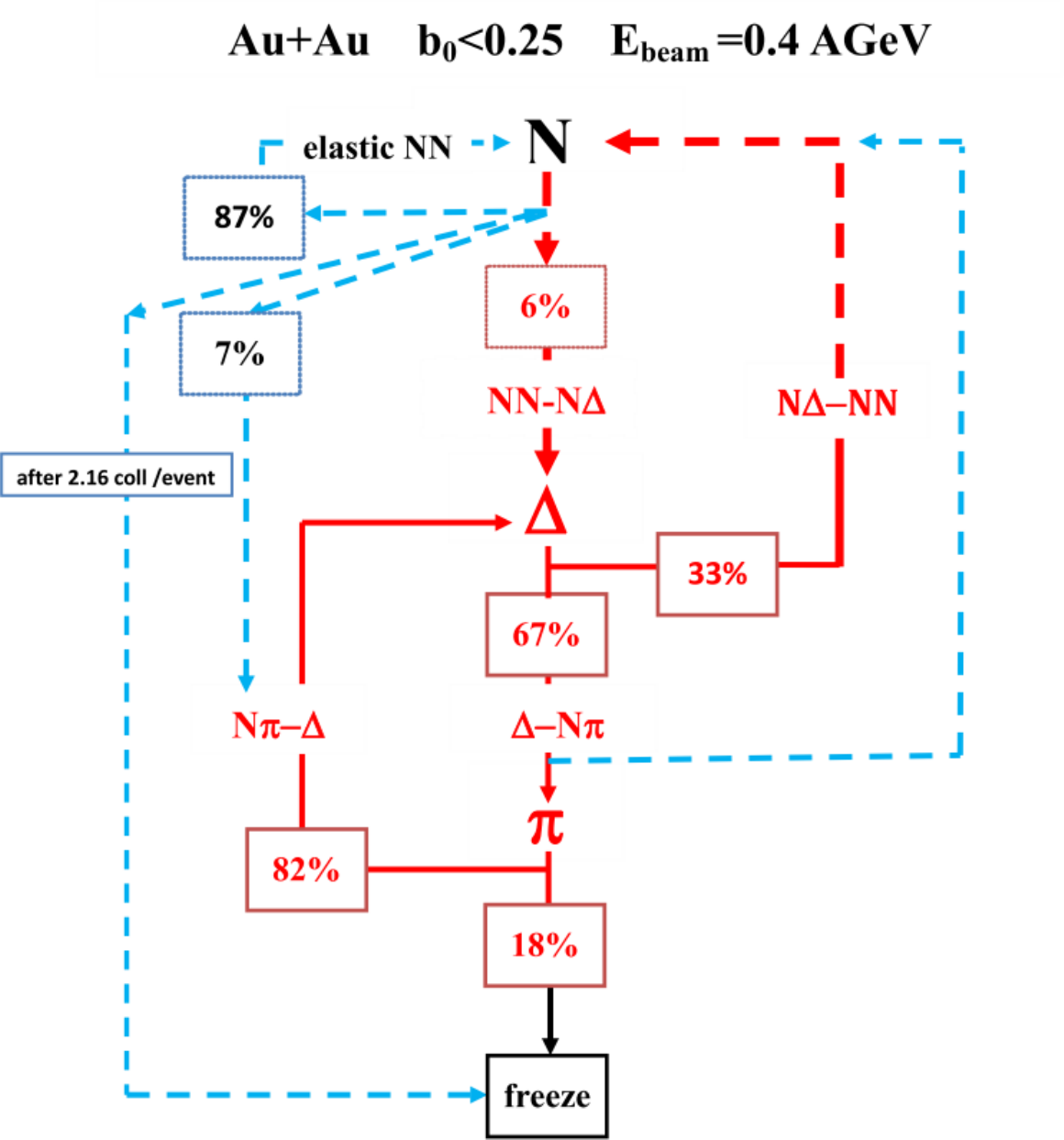}
\setlength{\abovecaptionskip}{0pt}
\vspace{2em}
\caption{(Color online) $N-\pi-\Delta$ loops in the UrQMD model, for Au+Au at $E_{beam}$=0.4A GeV and $b_0<0.25$ (More details see the text).}
\setlength{\belowcaptionskip}{0pt}
\label{fig:Ncoll-ratios}
\end{figure}

By integrating the collision/decay number over 0-60 fm/$c$, we can calculate the probability of different processes in $N-\pi-\Delta$ loops in the UrQMD model. In Figure~\ref{fig:Ncoll-ratios}, we plot the probability of different processes which are similar to the scheme plotted by Bass \textit{et al}. in Ref.~\cite{bass1995}. The reaction system is Au+Au at $E_{beam}$=0.4A GeV and reduced impact parameter $b_0<0.25$.  
As shown in the scheme, $\Delta$ resonances are initially produced via inelastic nucleon nucleon scattering, i.e., $NN\to N\Delta$, which is also observed from Figure~\ref{fig:contour} (b1) and Figure~\ref{fig:Ncoll-t}. For Au+Au collisions at 0.4A GeV, $\sim$87\% of total NN collisions are elastic collisions, and only $\sim$6\% belong to inelastic collisions of $NN\to N\Delta$. For $\Delta$s, there are two kinds of $\Delta$-loops, type (I) is the $NN\to N\Delta$ and $N\Delta\to NN$, and type (II) is the $\Delta\to N\pi$ and $N\pi\to \Delta$. Qualitatively, one can find that the 67\% of $\Delta$s decay into nucleon and pion, and 33\% of them participate in the $\Delta N\to NN$ collisions. Around 0.4A GeV, there are average 4.5 $\Delta-N-\pi$ loops before the pion freeze out. Based on the Fig.~\ref{fig:Ncoll-t} and scheme in Fig.~\ref{fig:Ncoll-ratios}, one can imagine that the type (I) $\Delta$-loop can keep the sensitivity of $\pi^-/\pi^+$ to the high density symmetry energy, but the type (II) $\Delta$-loop degrades the sensitivity of $\pi^-/\pi^+$ to high density symmetry energy.

In reality, both the type (I) and (II) loops make the sensitivity of $\pi^-/\pi^+$ to symmetry energy reduced resulting from the mutual change of the charge state of baryons in the $\Delta$ and $\pi$ production and absorption processes. For example, in the process of $n+n\to p+\Delta^-$, a neutron is converted to a proton in the simulations and it may change back to neutron during the evolution. It means the local symmetry potential will be changed. For $n+n\to n+\Delta^0$, a neutron changes to $\Delta^0$ which potential is the mixing of neutron and proton potential as in Eq.~(\ref{eq:VDelta}). The isospin and its third component of $i$th particle may change with the time, and thus the effects of symmetry energy on nucleons and $\Delta$ resonance are weakened, especially at high beam energies where the collision frequencies are large.

\subsection{Characteristic density of pion observable}
Considering the complicated process of pion production as discussed above, now an important question one has to answer is which density region is eventually probed by $M(\pi)$ and $\pi^-/\pi^+$, and we named this density region as `characteristic density'. There were some effort to answer this question by switching on/off the symmetry energy at different density region and check its influence on the pion multiplicity and its ratio~\cite{HLLiu15}. It seems a direct way, but the abrupt changes of the symmetry energy in the different density region could cause the unphysical force in the transport model simulations. Thus, it stimulate us to extract the characteristic density based on spatiotemporal evolution of pion production.

One method we proposed is to calculate the pion production rate weighted average density during the pion passing time in transport model simulations,
\begin{equation}
<\rho_{c}>_\pi=\frac{\int_{t_0}^{t_1} R_\pi(t)\rho_{c}(t) dt}{\int_{t_0}^{t_1} R_\pi(t)dt}.
\end{equation}
The $R_\pi(t)=\frac{dM_\pi(t)}{dt}$ is the pion production rate at certain time, and $\rho_{c}(t)$ is an averaged central density, which is calculated in a sphere with a radius equal to 3.35 fm centered at the center of mass of reaction system. The integral is from $t_0$=0 fm/$c$ to $t_1$=60 fm/$c$ in this work.

\begin{figure}[htbp]
\centering
\includegraphics[angle=0,scale=0.45]{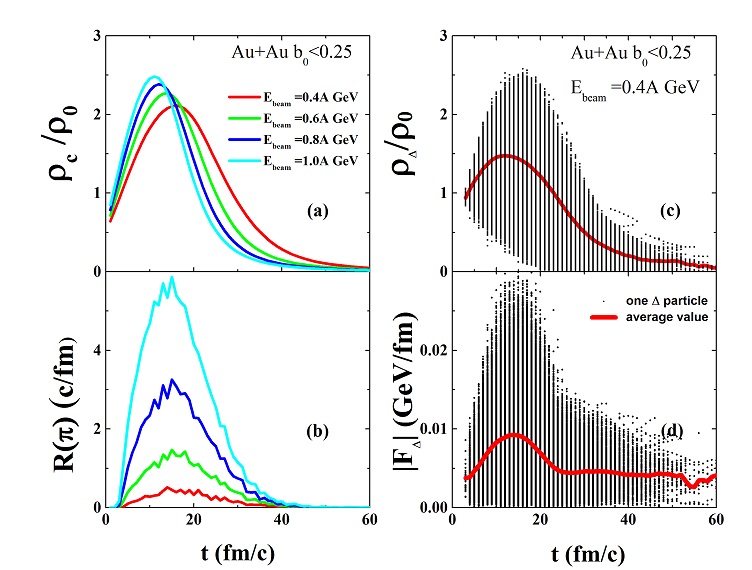}
\setlength{\abovecaptionskip}{0pt}
\vspace{2em}
\caption{(Color online) Time evolution of the averaged density in the center of reaction system (a), pion production rate (b). Panels (c) and (d), the density and force of $\Delta$ obtained from thousands events.}
\setlength{\belowcaptionskip}{0pt}
\label{fig:rhow}
\end{figure}

In panel (a) of Figure~\ref{fig:rhow}, we plot the time evolution of $\rho_c$ for Au+Au at the beam energy ranging from 0.4A GeV to 1.0A GeV with different colors. As one expected, the higher the beam energy is, the larger the compressed density is achieved, and the faster expansion is observed with time evolution. The panel (b) shows $R_\pi(t)$ as a function of time. In the beam energy region we studied, the $R_\pi(t)$ reaches maximum around 15 fm/$c$. At the beam energy of 0.4A GeV, the system expands to subnormal density after 28 fm/$c$ but pions are continuously produced till $\sim$40 fm/$c$. It means that the freeze out pions are not only from the high density region but also from the low density region at later stage. The same behaviors can be observed even at the beam energy of 1.0A GeV, therefor, it means that the subsequent interactions on the pion production tend to erase the effect of the initial decay that has taken place at high density.

\begin{figure}[htbp]
\centering
\includegraphics[angle=0,scale=0.32]{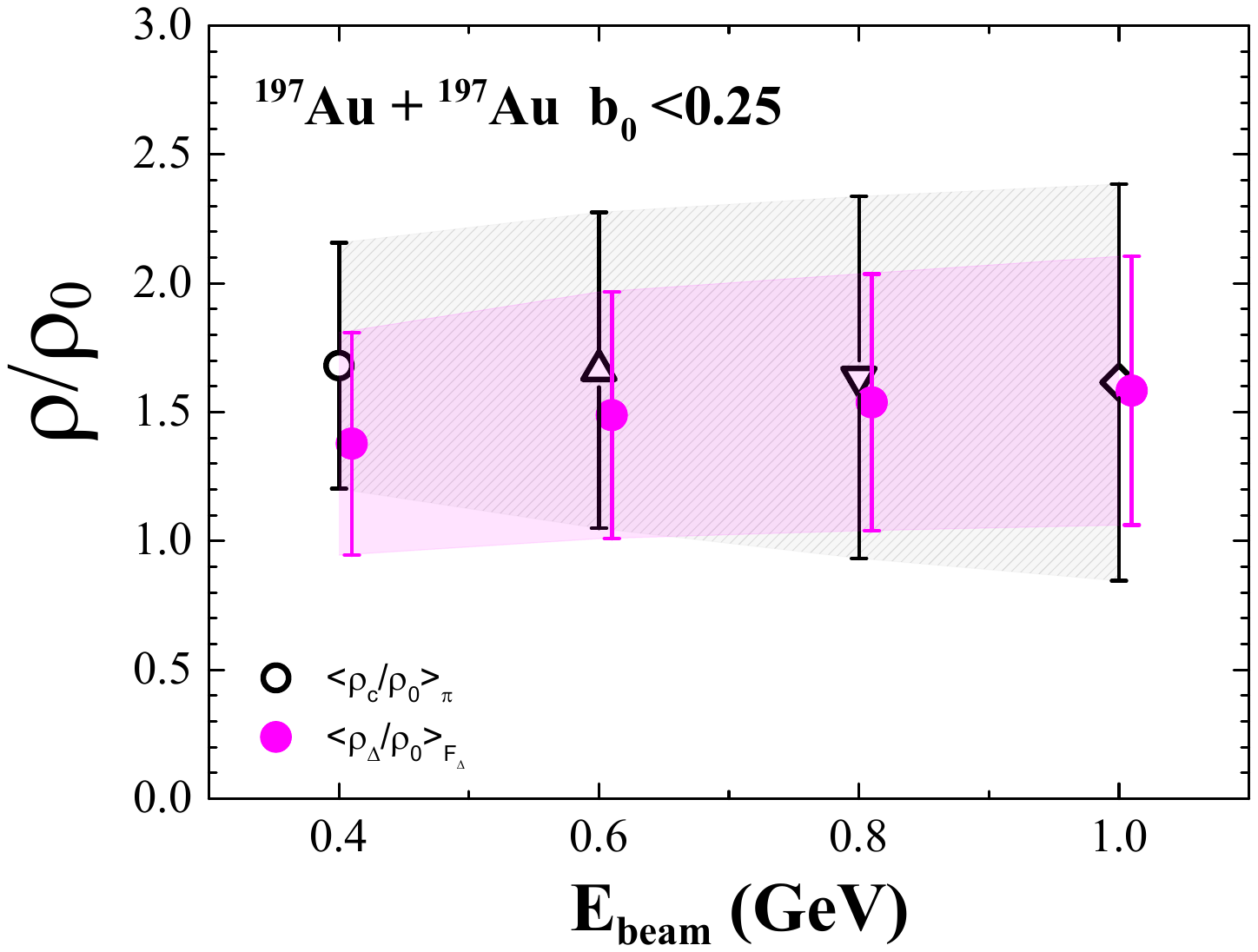}
\setlength{\abovecaptionskip}{0pt}
\vspace{2em}
\caption{(Color online) Pion-weighted density (black symbols) and the force acting on $\Delta$ weighted density (magenta symbols) at the beam energy from 0.4 - 1.0A GeV.}
\setlength{\belowcaptionskip}{0pt}
\label{fig:rhowpf}
\end{figure}

In Figure~\ref{fig:rhowpf}, the $<\rho_c>_\pi$ are presented and the values are around 1.6$\rho_0$ at all the beam energies we studied, which does not increase obviously with the beam energy increasing. It seems contradictory with the impression we have. Actually, this is because that the system spends a larger fraction of the total time till freeze-out at lower densities for higher beam energy than that for lower beam energy. For example, at the beam energy of 1A GeV, the system takes about 19 fm/$c$ at $\rho>\rho_0$, and about 41 fm/$c$ at $\rho<\rho_0$, but at the beam energy of 0.4A GeV, the system takes about 26 fm/$c$ at $\rho>\rho_0$, and about 34 fm/$c$ at $\rho<\rho_0$.
Thus, the weighted average central density which is obtained with pion production rate over time, becomes almost constant. Due to the above behaviors, the density variance which is defined as,
\begin{equation}
\sigma_\rho^2=\frac{\int_{t_0}^{t_1} R_\pi(t)(\rho_c(t)-<\rho_c>_{\pi})^2 dt}{\int_{t_0}^{t_1} R_\pi(t)dt},
\end{equation}
increases with the beam energy increasing.

As a comprehensive understanding of the characteristic density related to the pion productions, we also investigate $\rho^{(i)}_\Delta(t)$ and $|F^{(i)}_\Delta(t)|$ as a function of time for Au+Au at 0.4A GeV. The $\rho^{(i)}_\Delta(t)$ and $|F^{(i)}_\Delta(t)|$ are the density where the $i$th $\Delta$ is, and the force acting on $i$th $\Delta$, respectively, and they are plotted in panel (c) and (d) of Figure~\ref{fig:rhow}. The black points are the density or force obtained from different events, and one can observe $\rho_\Delta$ distributes from subnormal density to supranormal density even at the stage of highly compressed phase. It leads to the average values of $\rho_\Delta/\rho_0\le1.5$, as illustrated with the red line in panel (c). Based on the $\rho^{(i)}_\Delta$ and $|F^{(i)}_\Delta|$, we calculate the $\Delta$-force weighted density as follows,
\begin{equation}
<\rho>_{F_{\Delta}}=\frac{\int_{t_0}^{t_1} \sum_i |F^{(i)}_\Delta(t)|\rho^{(i)}_\Delta(t) dt}{\int_{t_0}^{t_1} \sum_i |F^{(i)}_\Delta(t)| dt}.
\end{equation}

The $|F_\Delta|$-weighted average density and its standard deviation are also presented in Figure~\ref{fig:rhowpf} with magenta symbols and shadows. Our calculations show $|F_\Delta|$-weighted average densities are slightly smaller than one obtained with $R(\pi)$-weighted density and have an increasing trend with the beam energy. It can be understood from Figure~\ref{fig:Ncoll-t} and Figure~\ref{fig:rhow}, where the $\Delta$s are found to exist shorter time than pions, and average $\rho_\Delta$ is smaller than the average central density of system. Even there is a little difference, both $|F_\Delta|$-weighted average density and $R_\pi$-weighted average density methods show the pion observable carry the information of compressed nuclear matter in the density region of 1-2.5 times normal density, which is larger than the force-weighted density for flow observable in 0.7-2.2 times normal density at the beam energy from 0.4A GeV to 1.0A GeV in Ref.~\cite{Fevre16}. Based on our calculations and results in Ref.~\cite{Fevre16}, for probing the symmetry energy $> 2.5\rho_0$ with HICs, one may need to measure, such as Kaon and $\Sigma$~\cite{QFL2005JPG31,QFL2005PRCsigma,Ferini05,Fuchs06}, or propose new probes.

\subsection{Effects of $\sigma^{OBEM}_{N\Delta\to NN}$, incompressibility $K_0$ and symmetry energy on the $M(\pi)$ and $\pi^-/\pi^+$}

\begin{figure}[htbp]
\centering
\includegraphics[angle=0,scale=0.43]{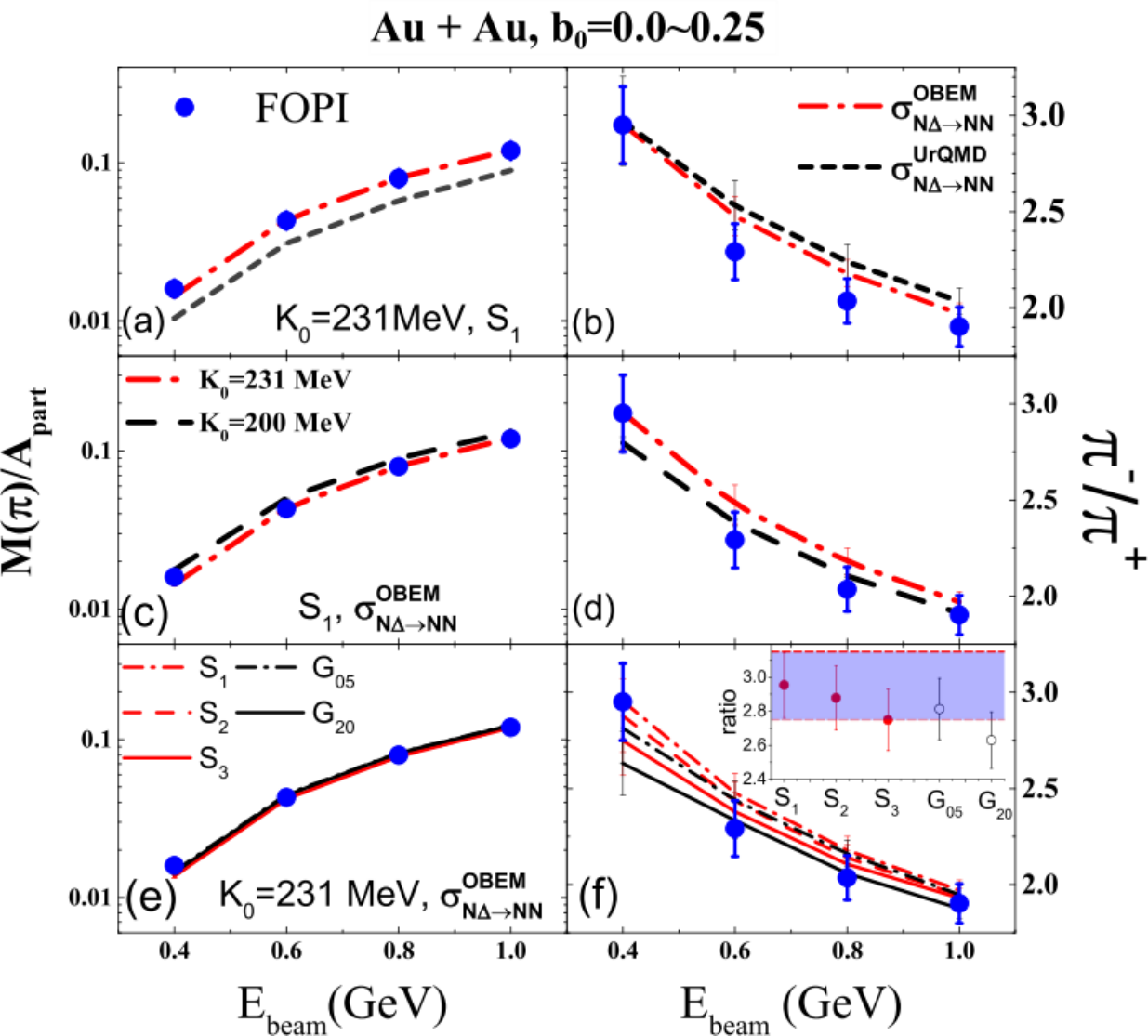}
\setlength{\abovecaptionskip}{0pt}
\vspace{2em}
\caption{(Color online) The excitation function of the $M(\pi)/A_{part}$ (left panels), and $\pi^-/\pi^+$ (right panels), for central collisions of $^{197}\textrm{Au}+^{197}\textrm{Au}$ reaction. (a) and (b) are the results with ${\sigma _{N\Delta \rightarrow NN}^{OBEM}}$ (red lines) and ${\sigma _{N\Delta \rightarrow NN}^{UrQMD}}$ (gray lines). (c) and (d) are the results with $K_0=200$ MeV (black lines) and $K_0=231$ MeV (red lines). (e) and (f) are results with different forms of symmetry energy. The FOPI data are shown as solid symbols ~\cite{Reisd07}.}
\setlength{\belowcaptionskip}{0pt}
\label{fig:piratio}
\end{figure}

Now, let's investigate the influece of $\sigma_{N\Delta\to NN}$, incompressibility $K_0$ and symmetry energy on the $M(\pi)$ and $\pi^-/\pi^+$. All simulations are performed with 200,000 events and impact parameters range from 0 to 3.35 fm. The upper panels of Figure~\ref{fig:piratio} are the results of the $M(\pi)/A_{part}$ and the $\pi^-/\pi^+$ ratio obtained with two kinds of $N\Delta\to NN$ cross section in the UrQMD calculations in case of $K_0=231$ MeV and symmetry energy with $S_1$, for Au+Au at beam energy from 0.4A GeV to 1.0A GeV. The gray lines are the results obtained with the form of $\sigma^{UrQMD}_{N\Delta\to NN}(\sqrt s)$ in UrQMD~\cite{QFL2006JPG32}, where the $M(\pi)$ is underestimated as that found in the Ref.~\cite{QFL2006JPG32}, and the $\pi^-/\pi^+$ ratios are overestimated. The red lines are the results obtained with $\sigma^{OBEM}_{N\Delta\to NN}(\sqrt s, m_\Delta)$~\cite{YCui2019} in UrQMD model. With the $\sigma_{N\Delta\to NN}^{OBEM}(\sqrt s, m_\Delta)$, the $M(\pi)$ is enhanced and $\pi^-/\pi^+$ are suppressed. Both the $M(\pi)$ and $\pi^-/\pi^+$ ratio are close to the FOPI data within the experimental uncertainties. The refined descriptions can be understood from Figure~\ref{fig:sigma}, where the value of $\sigma_{N\Delta\to NN}^{OBEM}(\sqrt s, m_\Delta)$ is lower than that of the $\sigma_{N\Delta\to NN}^{UrQMD}(\sqrt s)$ for low mass $\Delta$ near the threshold energy. As a result, it leads about 67\% of the produced $\Delta$s, which are produced by $NN\to N\Delta$ process, entering into the $\Delta\to N\pi$ process, while the probability was only about 51\% if the $\sigma_{N\Delta\to NN}^{UrQMD}(\sqrt s)$ used in the UrQMD calculations. Thus, the $\pi$ multiplicity are enhanced and $\pi^-/\pi^+$ ratios are decreased with $\sigma_{N\Delta\to NN}^{OBEM}(\sqrt s, m_\Delta)$ compared with that with $\sigma^{UrQMD}_{N\Delta\to NN}$.

The panel (c) and (d) show the influence of $K_0$ on the $M(\pi)$ and $\pi^-/\pi^+$ with the symmetry energy of $S_1$ and $\sigma_{N\Delta\to NN}^{OBEM}$ in the UrQMD calculations. The black lines are the results of $K_0=200$ MeV, and red lines are the results of $K_0=231$ MeV. The calculations showed that more pions were produced for the case of $K_0$=200 MeV, and the values of $M(\pi)$ were on the upper limits of the data uncertainties, while the $\pi^-/\pi^+$ ratios were reduced by less than 5\%.

To see the effects of density dependence of the symmetry energy on the $M(\pi)$ and $\pi^-/\pi^+$, we calculate central collisions for Au+Au with 5 kinds of density dependence of the symmetry energy, i.e., $S_1$, $S_2$, $S_3$, $G_{05}$, and $G_{20}$, 
as in Table.~\ref{tab:table2} with the $\sigma^{OBEM}_{N\Delta\to NN}(\sqrt s, m_\Delta)$ and $K_0=231$ MeV. The selected five forms of the symmetry energy include the uncertainties of the symmetry energy coefficient ($S_0$) and the slope of symmetry energy ($L$). All the results are plotted in the panel (e) and (f) of Figure~\ref{fig:piratio}. The lines with black color are the results obtained with $G_{05}$ and $G_{20}$, 
and red color are the results obtained with $S_1$, $S_2$, and $S_3$. As illustrated in bottom panels, the $M(\pi)$ is not sensitive to the density dependence of symmetry energy, while the $\pi^-/\pi^+$ ratio depends on the stiffness of symmetry energy, but the sensitivity to symmetry is weak.

As the discussions in previous, our results confirm again that the $\pi^-/\pi^+$ ratios of the HICs near the threshold energy is more sensitive to symmetry energy than that at higher beam energies.  To clearly see the effect of symmetry energy, we also show the calculated results of $\pi^-/\pi^+$ at 0.4A GeV in the inset of Figure~\ref{fig:piratio}. One can see that $\pi^-/\pi^+$ ratios do not simply decrease with the $L$ increasing, where the effect of $S_0$ is also important as well as $L$. For example, by comparing the results of $S_2$ ($S_0$=34 MeV, $L$=81 MeV) and $G_{05}$ ($S_0$=32.5 MeV, $L$=54 MeV), one can find that the $\pi^-/\pi^+$ of $S_2$ is greater than that of $G_{05}$, where the $L$ value of $S_2$ is greater than that of $G_{05}$. For $S_1$ and $G_{05}$, their $L$ values are close, but the difference of $\pi^-/\pi^+$ between two symmetry energy forms is close to the difference with $S_1$ and $S_3$ where the $L$=46 MeV and $L$=104 MeV. Those calculations show that the calculations with more parameter sets, especially in mutli-dimensional parameter space, could help us build the correlation relationship between $S_0$ and $L$ by describing the pion observables.

\subsection{Symmetry energy from $\pi^-/\pi^+$ ratio and its model dependence}

As in many previous work, we tried to compare our calculations to the FOPI data and learn a model dependent information of symmetry energy at 1-2 times normal density. The calculations are performed with the parameter sets $K_0=231$ MeV and $\sigma^{OBEM}_{N\Delta\rightarrow NN}$, and five kinds of symmetry energy, i.e., $S_1$, $S_2$, $S_3$, $G_{05}$ and $G_{20}$. In Table~\ref{tab:table3}, $\chi_i^2=\frac{(Y_i^{th}-Y_i^{exp})^2}{\sigma_{i,r}^2}$ are presented for different beam energies, where $Y_i^{th}$ is theoretical result of $\pi^-/\pi^+$, and $Y_i^{exp}$ and $\sigma_{i,r}$ are the data and its uncertainty, respectively.

At the beam energy of 1.0A GeV, the $\chi_i^2$ values for all kinds of symmetry energy we used are less than 1, and it means we can not rule out any kinds of symmetry energy we used with the 1.0A GeV data even within 1$\sigma$ uncertainties. With the beam energy decreasing, for example, the $\chi_i^2$ values of $S_1$ and $G_{05}$ at 0.6 and 0.8A GeV are greater than 1, and it means the $S_1$ and $G_{05}$ are ruled out within 1$\sigma$ uncertainties. At the beam energy of 0.4A GeV, the results with $G_{05}$, $S_1$, $S_2$, and $S_3$ fall into the data uncertainties, but for the results of $G_{20}$, only their upper limit falls into the data region and its $\chi_i^2$ is up to 2.56. Comparing the $\chi^2_i$ with that obtained at higher beam energies, the sensitivity of $\pi^-/\pi^+$ to symmetry energy at 0.4A GeV becomes a little large. It hints the experiments at the beam energy lower than 0.4 GeV per nucleon may be a better choice to enhance the symmetry energy effects. The different behaviors of $\chi^2$ varying with symmetry energy implies that a uniform description of the pion data at the beam energy ranging from 0.4A GeV to 1.0A GeV still need more works from the sides of transport models. 

\begin{table}[htbp]
\caption{\label{tab:table3}%
Entries in 2th-5th row are the $\chi_i^2=\frac{(Y_i^{th}-Y_i^{exp})^2}{\sigma_{i,r}^2}$ of $\pi^-/\pi^+$ ratios for 5 kinds of symmetry energy. Last row is the corresponding $\chi^2$ values.}
\centering
\begin{tabular}{lccccc}
\hline
\hline
$E_{beam}$ & $S_1$ & $S_2$ & $S_3$ & $G_{05}$ & $G_{20}$ \\
\hline
0.4 & 0.00 & 0.13 & 1.01 & 0.47 & 2.56 \\
0.6 & 1.61 & 0.96 & 0.36 & 1.03 & 0.08 \\
0.8 & 1.55 & 0.80 & 0.38 & 1.18 & 0.03 \\
1.0 & 0.44 & 0.17 & 0.07 & 0.20 & 0.06 \\
\hline
\hline
\end{tabular}
\end{table}

\begin{figure}[htbp]
\centering
\includegraphics[scale=0.35]{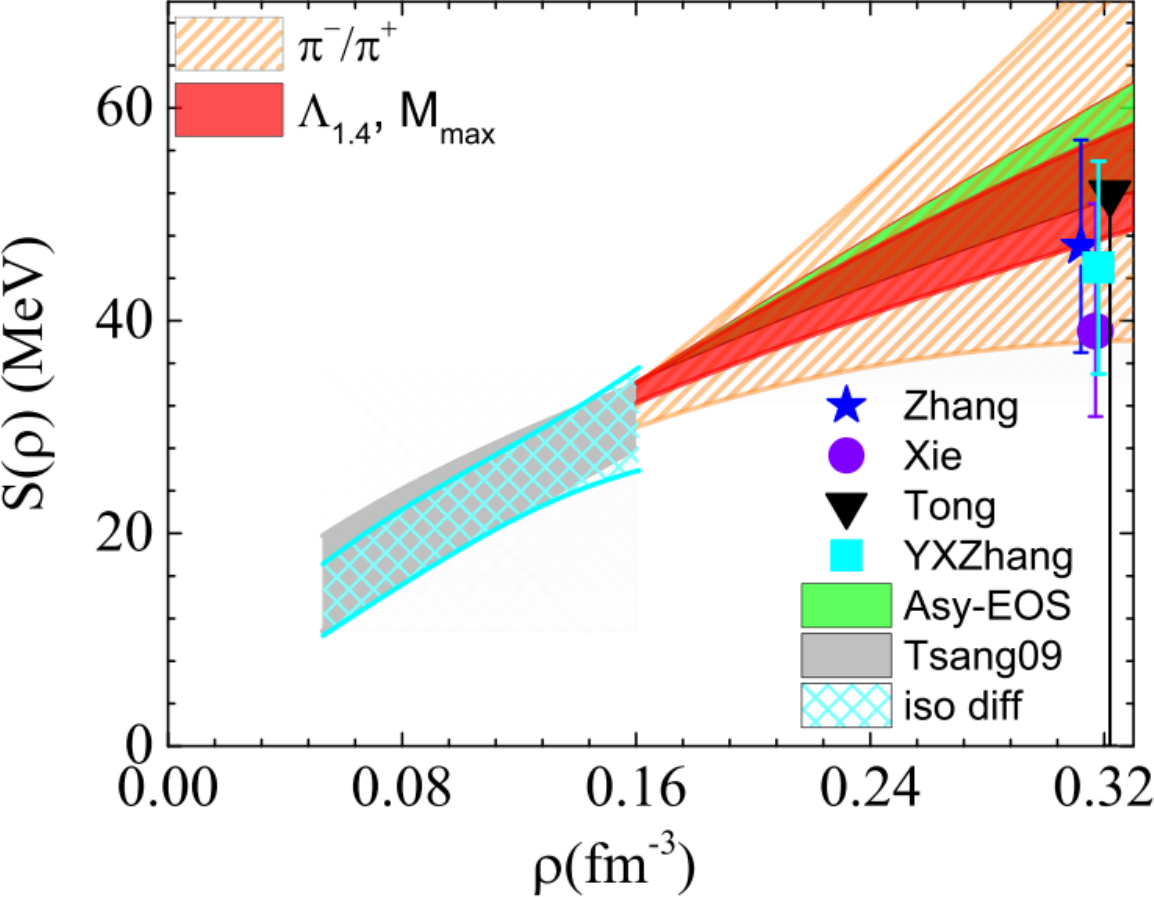}
\caption{(Color online)
Constraints of density dependence of the symmetry energy from $\pi^-/\pi^+$ in UrQMD (orange shaded region). Red solid region is the constraints from $\pi^-/\pi^+$, $\Lambda_{1.4}$, and $M_{max}$ (see more details in the text).}
\setlength{\belowcaptionskip}{0pt}
\label{fig:srho-cons}
\end{figure}

Based on the above comparisons and favored sets at the beam energy of 0.4A GeV, we indirectly obtain the corresponding density dependence of the symmetry energy for cold nuclear matter, which is shown in Figure~\ref{fig:srho-cons}. The orange shaded region is the constraint obtained from $\pi^-/\pi^+$ in this work within 1$\sigma$ uncertainty. The light green region is the constraint from ASY-EOS flow data~\cite{Russotto2016PRC}, which is a narrow band since it fixed the symmetry energy coefficients $S_0=34$ MeV. The blue star, purple circle and black triangle are the constraints of symmetry energy at $2\rho_0$ from neutron star analysis, $S(2\rho_0)=47 \pm10$MeV~\cite{NBZhang19}, $39\pm^{12}_8$ MeV~\cite{WJXie19}, $\le 53$ MeV~\cite{HTong19}, respectively. The cyan shaded region and square are the constraint from the combination analysis of isospin diffusion data, neutron skin, and neutron stars~\cite{Zhang20} in five-dimensional parameter space, which predict $S(2\rho_0)$=35-55 MeV. The gray shaded region is the constraints of symmetry energy at subsaturation density based on the heavy ion collision observables, such as isospin diffusion, isospin transport ratio as a function of rapidity~\cite{Tsang09,Zhang20}. As observed in Figure~\ref{fig:srho-cons}, the constraints on the symmetry energy at high density is consistent with the constraints from flow data and neutron stars, but with large uncertainties.

However, the other factors may also influence the prediction of $M(\pi)$ and $\pi^-/\pi^+$ in some extent, such as momentum dependent interaction~\cite{QFL2006JPG32}, symmetry potential of $\Delta$(1232)~\cite{BALi15,Cozma2016PLB}, the in-medium threshold effect~\cite{SK2015PRC}, and pion potential~\cite{ZZhang17,JH2014PRC}, and thus on the exact values of the constrained symmetry energy.

For example, the momentum dependent interaction also play important roles. In our previous discussions, the parameters of $t_4$ and $t_5$ in the MDI are determined by fitting $p+Ca$ data of Arnold~\cite{Arnold82} which yield the effective mass $m^*/m=0.77$ at Fermi momentum. After the analysis of a wealth of data by Hama~\cite{Hama90}, this form has been updated~\cite{Hartnack94} which predict a smaller effective mass. We also tried to adjust the values of $t_4$ and $t_5$ in the MDI form we used, to refit the Hama's data within the range of $E_{inc}$=1 GeV and keep $K_0=231$ MeV. If we take $t_4=3.05$ MeV and $t_5=5\times 10^{-4}$ MeV, which predict the $m^*/m=0.635$ and can well reproduce the data within $E_{inc}$=0.7 GeV. With this updated MDI sets, the UrQMD calculations show a non-negligible amount on the pion multiplicity. For example, the pion multiplicities are reduced by about 30\%, and $\pi^-/\pi^+$ ratios are reduced by $<$15\% when we take $S_1$ parameter sets. Thus, to describe the FOPI data, one may need to consider the medium correction on the Delta and pion related cross sections, or a much softer symmetry energy. Especially, related to $\pi^-/\pi^+$ ratios, the isospin dependent MDI should be carefully investigated but it is still an open question.


For the symmetry potential of $\Delta$, even its strength is not very clear, the recent calculations~\cite{Cozma2016PLB,BALi15} show that the standard $\Delta$ potential of Eq.(\ref{eq:VDelta}) is a reasonable choice. When we artificially enhance the strength of $\Delta$ symmetry potential to 2 times than that in Eq.~(\ref{eq:VDelta}), the calculations with UrQMD also show negligible effect. It is consistent with the studies by B. A. Li~\cite{BALi15}, where they have shown that the total and differential $\pi^-/\pi^+$ ratio in heavy-ion collisions above the threshold energy are weakly influenced by the completely unknown symmetry (isovector) potential of the $\Delta$(1232) resonance, owing to the very short lifetimes of $\Delta$ resonances~\cite{BALi15}. Although the results from the Tubingen QMD model showed that the constraint of symmetry energy extracted from $\pi^-/\pi^+$ was highly sensitive to the strength of the iso-vector $\Delta$ potential~\cite{Cozma2016PLB}, they have also found that the standard $\Delta$ potential of Eq.(\ref{eq:VDelta}) is suitable if the constrained $L$ values are consistent with the results from nuclear structure and reaction studies.

For the in-medium threshold effect, the RVUU model calculations show that the in-medium threshold effect enhances both the $M(\pi)$ and $\pi^-/\pi^+$ compared to those without this effect~\cite{SK2015PRC}. However, the calculations with RVUU model show that including the pion potential decreases the $\pi^-/\pi^+$ ratio. Thus, including both the threshold effect and pion potential will lead to a cancelled effect on $\pi^-/\pi^+$. To describe the experimental data, while a softer symmetry energy with the slope parameter $L$ = 59 MeV is needed in RVUU calculations~\cite{ZZhang17}. Furthermore, including the threshold effects and pion potential in transport models not only need to change the threshold energy, but also require to treat the in-medium cross sections of $NN\leftrightarrows N\Delta$ and $N\pi\leftrightarrows \Delta$\cite{ZZhang17,Cui18,Cui20} consistently. A difficulty on treating the detailed balance of in-medium $\Delta\rightleftharpoons N\pi$ in transport models~\cite{ZZhang19} also need to overcome in future.

The model dependence of the constraints of symmetry energy via $\pi^-/\pi^+$ stimulate us to do combinatory analysis of multi-observables in HICs and properties of neutron stars, such as pion yields and ratios, neutron proton flow, neutron to proton yield ratios, mass-radius relationship and tidal deformability of neutron stars, and so on, to reduce the uncertainties in future, which naturally require more experimental data near the threshold energy and a Bayesian analysis in multi-dimensional parameter space. 



\subsection{Constraints from neutron stars}
By using the interactions used in this work, we calculate the equation of state (EoS) of neutron star matter in the density range $0.5\rho_0<\rho<3\rho_0$ which is obtained by simultaneously fulfilling the $\beta$-stability and local charge neutrality conditions, including the contributions of $e^-$ and $\mu^-$. At subsaturation densities, the pasta phases of nuclear matter emerge, we thus adopt the EoSs presented in Refs.~~\cite{Feynman49, Baym71, Negele73} at $\rho<0.08\ \mathrm{fm}^{-3}$. For the density region above $3\rho_0$, the UrQMD density functional does not apply and we adopt a polytropic EoS~~\cite{Lattimer16,Fattoyev13,CYTsang19}, where the pressure is given by $P =\kappa \rho^{\gamma'}$. At given $\gamma'$, the parameter $\kappa$ and energy density are fixed according to the continuity condition of pressure and baryon chemical potential at $\rho=3\rho_0$.

\begin{figure}[htbp]
\centering
\includegraphics[scale=0.35]{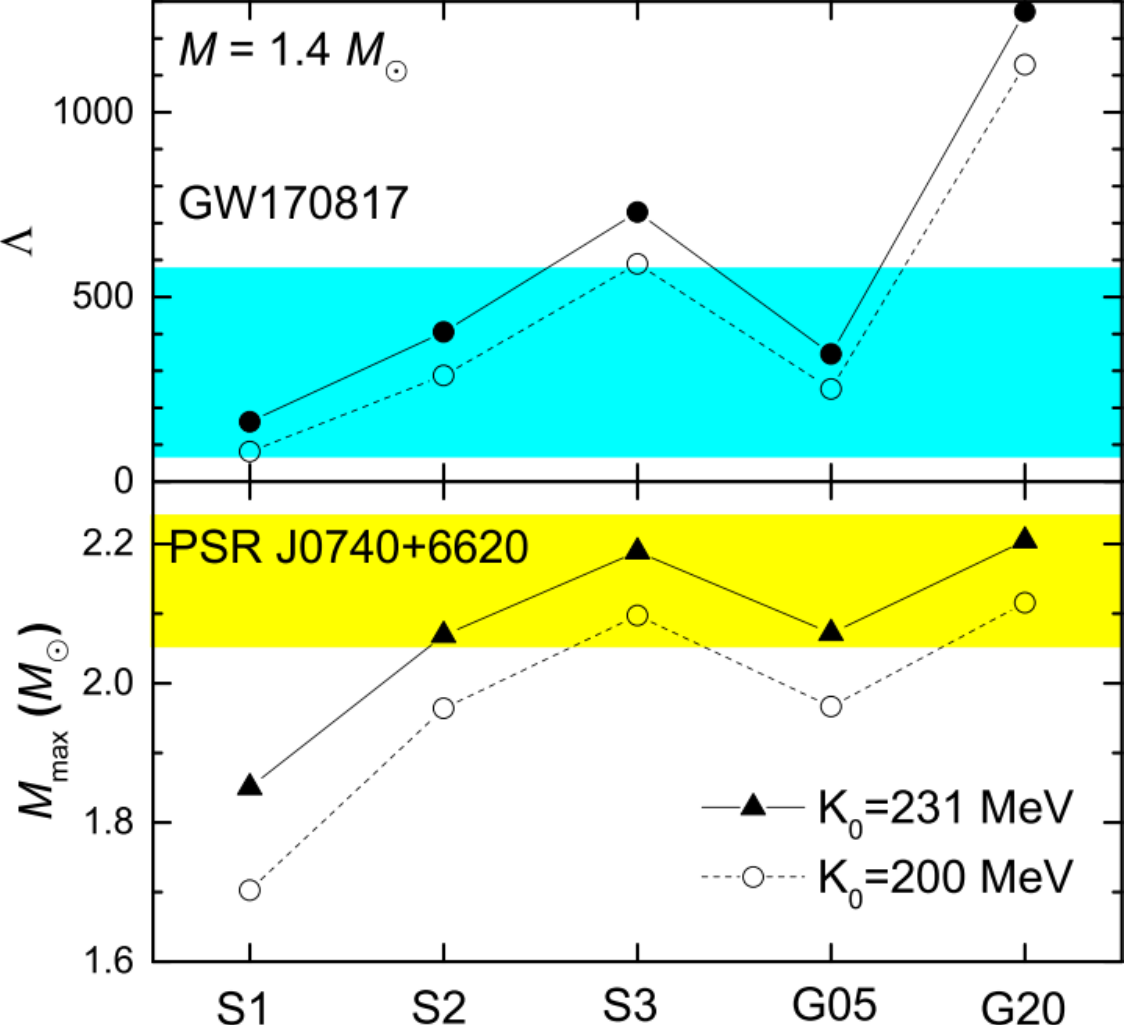}
\caption{(Color online) Tidal deformability and maximum mass of neutron star obtained with the interaction we used in UrQMD model. The shaded regions are the constrained values of $\Lambda_{1.4}$ and $M_{max}$~\cite{Abbott18,Cromartie19}.}
\setlength{\belowcaptionskip}{0pt}
\label{fig:mr-ns}
\end{figure}

The structure of a neutron star is then obtained by solving the Tolman-Oppenheimer-Volkov equation, while the tidal deformability is estimated with $\Lambda = \frac{2 k_2}{3}\left( \frac{R}{G M} \right)^5$~~\cite{Damour09, Hinderer10,Postnikov10}. In Figure~\ref{fig:mr-ns}, we present the obtained tidal deformability at $M=1.4\ M_\odot$ and the maximum mass based on those parameter sets we used. The solid symbols are the results obtained with $K_0=231$ MeV and open symbols are the results obtained with $K_0=200$ MeV. The shaded region is the constraints on $\Lambda$ and $M_{max}$~\cite{Abbott18,Cromartie19}, obtained with the binary neutron star mergering event GW170817 ($70\leq\Lambda_{1.4}\leq 580$)~\cite{Abbott18} and the observational mass of PSR J0740+6620 ($2.14{}_{-0.09}^{+0.10}\ M_\odot$)~\cite{Cromartie19} without violating the casuality limit ($\gamma'\leq 2.9$).
In Table~\ref{tab:table4}, we present the $\chi^2_i$ of different forms of symmetry energy. For the case of $K_0=200$ MeV, it required the symmetry energy $G_{20}$ and $S_3$ for describing the observational mass of PSR J0740+6620 ($2.14{}_{-0.10}^{+0.09}\ M_\odot$)~~\cite{Cromartie19} and can not describe the data of $\Lambda$. Furthermore, the sets of $G_{20}$ and $S_3$ had a large slope of symmetry energy and were not consistent with the recent commonly accepted $L$ value~\cite{Latti13,BALi13,Tsang12,YanZhang}. If the $K_0=231$ MeV are adopted in the calculations, we finally find that the parameter sets $S_2$ and $G_{05}$ can reproduce the $\pi^-/\pi^+$, $\Lambda_{1.4}$ and $M_{max}$ simultaneously. The red solid region in Figure~\ref{fig:srho-cons} is the density dependence of the symmetry energy between $G_{05}$ and $S_2$. At two times normal density, the $S(2\rho_0)$ is in 48-58 MeV. This value is consistent with those obtained from ASY-EOS flow~\cite{Russotto2016PRC} and neutron star analysis~\cite{NBZhang19,WJXie19,HTong19,Zhang20} within their uncertainties. The corresponding radius of a 1.4 solar mass neutron star is also obtained with $12.0 \leq R_{1.4} \leq 12.5$ km.

\begin{table}[htbp]
\caption{\label{tab:table4}%
Entries in 2th, 3th, 5th, 6th row are the $\chi_i^2=\frac{(Y_i^{th}-Y_i^{exp})^2}{\sigma_{i,r}^2}$ of $M_{max}$ and $\Lambda$ for 5 kinds of symmetry energy. }
\centering
\begin{tabular}{lcccccc}
\hline
\hline
$K_0$ & Exp. & $S_1$ & $S_2$ & $S_3$ & $G_{05}$ & $G_{20}$ \\
\hline
200 & $M_{max}$ & 19.15 & 3.10 & 0.18 & 3.02 & 0.06 \\
 & $\Lambda$ & 0.07 & 0.06 &  1.05 & 0.02 & 5.79 \\
\hline
231 & $M_{max}$ & 8.38 & 0.51 & 0.24 & 0.46 & 0.42 \\
 & $\Lambda$ & 0.00 & 0.30 & 1.91 & 0.15 & 7.70 \\
\hline
\end{tabular}
\end{table}

\section{Summary and outlook}
In the framework of the UrQMD model, we analyze the pion production mechanism at the beam energy ranging from 0.4A GeV to 1.0A GeV. We demonstrate again that the pions experience averaged 4.5 times loops before freezing out. The loop of $N\pi\leftrightarrow\Delta$ and $NN\leftrightarrow N\Delta$ weaken the symmetry energy effect, especially at the beam energy above 0.6A GeV. Furthermore, we also analyze the character density probed by pion multiplicities and its ratios, and we find that the pion observables probe the symmetry energy in 1-2.5 normal density for beam energies ranges from 0.4 to 1.0A GeV.

With the $\sigma^{OBEM}_{N\Delta\to NN}(\sqrt s, m_\Delta)$, which take into account the $\Delta$-mass dependence of the M-matrix and $p_{N\Delta}(m_\Delta)$, in the UrQMD model calculations, the values of $M(\pi)$ are obviously enhanced and $\pi^-/\pi^+$ ratios are suppressed a little bit, and both the $M(\pi)$ and $\pi^-/\pi^+$ ratios are close to the FOPI data. By investigating the influence of symmetry energy on the $\pi^-/\pi^+$ and comparing the calculations to the FOPI data, we find that the parameter sets with symmetry energy at two times normal density $S(2\rho_0)=38-73$ MeV and the slope of symmetry energy $L=46-104$ MeV can describe the data within the data uncertainties. With the constraints from neutron star, such as $\Lambda_{1.4}$ and $M_{max}$, we obtain $S(2\rho_0)=48-58$ MeV and $L=54-81$ MeV, which is consistent with that from ASY-EOS flow data and neutron star within their uncertainties.

However, it is still a model dependent results. Especially, our calculations show the momentum dependent interaction has a non-negligible effect on the pion multiplicity. It naturally require multi-observables to reduce the model dependence. For example, simultaneously describing the nucleonic flow and pion ratios observables could be a better way to reduce the uncertainties of the constraints on symmetry energy at 1-2 times normal density. Another, the threshold effect and pion potential are also important for the energy spectral of pion production and ratios at subthreshold energy, which should be figure out in the future with the coming of new data~\cite{spirit}. It also requires to carefully develop the $\Delta$ and $\pi$ potential related issues, such as energy conservation and detailed balance, in the transport models. For extracting the symmetry energy at density above $2.5\rho_0$ in laboratory, our calculations show that we may need another probes, such as kaon or other new observables.

\section*{Acknowledgements}
This work was supported by the National Natural Science Foundation of China Nos. 11790325, 11875323, 11875125, 11947410
, 11705163, 11790320, 11790323, and 11961141003, the National Key R\&D Program of China under Grant No. 2018 YFA0404404, the Continuous Basic Scientific Research Project (No. WDJC-2019-13) and the funding of China Institute of Atomic Energy. Y.X. Zhang thanks Prof. H. Wolter, M. B. Tsang, A. Ono for the helpful discussions. The fruitful discussions with Jan Steinheimer are greatly appreciated. Q. Li also thanks the support of the ``Ten-Thousand Talent Program" of Zhejiang province. Y. Chen also thanks the partly support of the Chinese-Polish Joint project by National Science Foundation of China No.11961131010. We acknowledge support by the computing server C3S2 in Huzhou University.

\end{document}